\documentclass[doublecol]{epl2} 

\usepackage{graphicx}
\usepackage{soul}
\usepackage{color}

\newcommand{\be}{\begin{equation}}
\newcommand{\ee}{\end{equation}}
\newcommand{\bea}{\begin{eqnarray}}
\newcommand{\eea}{\end{eqnarray}}

\title{Signature of curved QFT effects on the optical properties of deformed graphene: Curved QFT effects on the optical properties of deformed graphene} 

\author{O. Oliveira\inst{1,2}, A. J. Chaves\inst{1}, W. de Paula\inst{1} \and T. Frederico\inst{1}}
\institute{                    
  \inst{1} Instituto Tecnologico de Aeronautica - Dep. de Fisica, DCTA, 12228-900 S\~ao Jos\'e dos Campos, Brazil\\
  \inst{2} Universidade de Coimbra - Dep. de Fisica, 3004-516 Coimbra, Portugal
  }
\pacs{}{}

\abstract{
The Dirac equation in curved space is used to  study
the optical transmittance of deformed graphene along
a given direction. Our theoretical analysis of the available experimental data for the light transmittance suggests that
the periodic ripple associated with the out-of-plane deformation observed in unstrained graphene explains the observations. 
Furthermore, the experimental uniaxial strained graphene for light transmittance show two features, namely
the modification of the $\cos^2\theta$ law  and the decrease of the amplitude of the oscillations with the polarization angle $\theta$,
which can be well accommodated within the theoretical analysis used here and provide further evidence of the
validity of using QFT in curved space to understand two dimensional materials.}

\begin{document}

\maketitle

\section{Introduction and Motivation}

Graphene is a two dimensional material which has been fabricated for the first time about ten years ago~\cite{Novoselov2004,Novoselov2005}.
Its electronic and optical properties are, typically, described assuming that the electrons live on a 2D flat surface
and using tight-binding models or, equivalently, the Dirac 
equation~\cite{Neto2009,Vozmediano2010,Peres2010,Chaves2011,Oliveira2011,Popovici2012,Cordeiro2013}.
However, the experimental evidence that the graphene sheet is not a planar surface is known
since the early times of graphene production~\cite{Morozov2006,Meyer2007,Ishigami2007,Stolyarova2007,Tapaszto2012}.
Furthermore, \textit{ab-initio} calculations~\cite{Fasolino2007} show that thermal fluctuations induce sizeable ripples in the graphene sheet.

From the point of view of the Dirac equation, deviations of graphene sheet from a planar surface can be considered, at least partially, via 
the formulation of Quantum Field Theory on curved  spaces; see e.g.~\cite{Vozmediano2010,GuineaNature2010,Levy2010,Juan2011,Chaves2014,Jakubsky2014} and references therein.
For very  large
geometric deformations, the overlap between the various orbitals is changed significantly. New types of terms should be 
added to the Dirac hamiltonian in order to take into account this effect in the material electronic structure~\cite{Juan2012}.
For small deformations, the usual free massless Dirac hamiltonian is the lowest order approximation to the Dirac equation in a curved space and,
hopefully, its geometric deformation can be handle by the introduction of curvature in the usual way and without appealing to new terms 
 in the hamiltonian.

Herein, we will investigate the optical properties of deformed (non-flat) graphene using the Dirac equation in curved space. As discussed previously,
one expects that such an approach is able to describe small deformations of the graphene sheet. This is not the only possible approach and, indeed, 
the effect of deforming or straining 
have also been investigated within the tight-binding model~\cite{Neto2009,Juan2011,Juan2012}. 
One can start from the tight-binding model with curvature and derive a continuum hamiltonian
taking into account the deformations of the material~\cite{Pereira2009,Winkler2010,Pellegrino2011,Linnik2012,Juan2012}.

In the approach based on the Dirac equation in a geometrically deformed space, the effect of
curvature can be rewritten in terms of pseudo gauge fields, whose physical origin can
be traced back to strain (in-plane deformations) and/or curvature (out-of-plane deformations)
of the pristine material, see e.g.~\cite{Chaves2014}.
These pseudo gauge fields can induce a zero-field quantum Hall effect~\cite{GuineaNature2010}, which
was observed experimentaly~\cite{Levy2010,Lu2012}. According to the authors, the observed pseudo magnetic fields can be as large as 
300--600 tesla. 

The geometric deformations of the graphene sheet, being it associated with strain or curvature, change the electronic and optical properties 
of the pristine material; see e.g~\cite{Vozmediano2008,Guinea2008,Isacsson2008,Pereira2009prl,Guinea2010,Pereira2010,Pellegrino2010,
Zhang2011,Kerner2012,Yan2013,Chaves2014, OlivaLeyva2014,Nguyen2014} for theoretical predictions
and~\cite{Morozov2006,Meyer2007,Ishigami2007,Stolyarova2007,Scniepp2008,Xu2009, Bao2009, Wang2011,Tapaszto2012,Koch2012,Bai2014,Ni2014,Park2014,
Plantey2014,Dong2014}  for experimental work.
From the experimental point of view, the realisation of such systems require the control of both the induced strain and the curvature 
of the graphene sheet. Strain and curvature appear as another tool to engineer the electronic and optical properties of graphene.
Further, the considerations discussed here with graphene also apply to any material whose theoretical understanding relies 
on the same principles as, for example, monolayers of transition metal dichalcogenides \cite{Casillas2014}.

In the present work, we will study the optical transmittance of deformed graphene from the point of view of a Dirac equation in curved space
and for a (small) deformation along a single direction based on our previous publication ~\cite{Chaves2014}. The present study was motivated 
by the recent experimental work of Ref.~\cite{Ni2014}, which measured the  optical conductivity in strained graphene using a laser with a frequency of  632.8 nm. Such a wave length
corresponds to red light, i.e. in the visible region. For our theoretical analysis, the experimental results corresponds to 
the infinite frequency limit of the formula derived in~\cite{Chaves2014}.
In the infinite frequency limit, it turns out that the optical conductivity is saturated by the interband electronic transitions.
The information about the geometric deformation of the graphene sheet is resumed in the arc length ($\xi$) of the
graphene sheet along the direction of the deformation and in a function ($F$) which ``measures'' the possible
electronic transitions from the valence to the conduction band.
For completeness, we will reproduce the main results of~\cite{Chaves2014} such that the reader can follow 
the discussion and have a better insight into our results and comparison to the data~\cite{Ni2014}.

This work is organized as follows. 
First we review the computation of the optical conductivity for deformed graphene. The special case of the small deformation
 is investigated in detail and general expressions, valid for any type of deformation, are derived. Furthermore, we discuss the outcome
 for the optical conductivity, in the limit of small deformations, for three different geometrical profiles  which are
 used to interpret  the experimental results~\cite{Ni2014}. Finally we compute the optical transmittance and
 discuss the available experimental results.

\section{The theoretical framework and the optical conductivity \label{Sec:Theo}}

In the following we will assume that the  graphene layer lies on the $xOy$ plane, i.e. its surface is defined by the
equation $z = 0$. If the graphene sheet is deformed in some way, its surface
is given by $z = h(x,y)$.  
Instead of studying a general two dimensional deformation, which includes strain 
and curvature, we will consider a periodic deformation along a single direction given by $z = h(x)= h\left(x+\frac{2\pi}{\omega}\right)$. 
Further, we will assume that the electronic wave functions satisfy periodic boundary conditions.
Recall that for sufficiently large samples, the boundary conditions have no influence on the electron physical properties.
The spatial periodicity can be seen either as having a regular structure in the deformation, which repeats every
$ 2 \pi / \omega$, or to take this distance as the full length of the graphene sample.
Such type of deformations were experimentally realised in~\cite{Bai2014,Ni2014} and, in particular, in~\cite{Ni2014} the
optical transmittance was measured for various deformations of the graphene.

We have shown elsewhere \cite{Chaves2014} that, by using the Dirac equation with a  periodic deformation along the $x$ axis and  within the 
linear response theory, the graphene optical conductivity in the long wavelength limit for the $x$ direction, $\Omega\to\infty $, is given by:
\be
   \frac{\Re \sigma^{\mathrm{inter}}_{xx}}{\sigma_0} \longrightarrow
    \left( \frac{2 \pi}{\xi} \right)^2 +  \frac{2}{\xi^2} \sum^{\infty}_{n = 1} \left| F(n) \right|^2 \label{f012}
\ee
where
\be
F(n) = \int_0^{2\pi}du \, e^{ i \phi(n,u)} ,  \label{f1}
\ee
and
\be
\phi(n,u)=\frac{2\pi n}{\xi}\int^u_0 dv \sqrt{ 1 + h^2_x(v )} \label{f2} \ .
\ee
The function $F(n)$ summarizes  the information on the allowed electronic transitions, together with
the geometry of the graphene sheet. Note that the dependence of the conductivity on the graphene
geometry comes via $h_x = d h / d x$ and not 
the profile function $h(x)$, i.e. the optical conductivity is sensitive to how the profile changes rather than 
to the profile itself. Moreover, it is the arc length $\xi=\int^{2\pi}_0 d v ~ \sqrt{ 1 + h^2_x (v) } \ ,$ which sets the scale for the optical conductivity.

The optical conductivity associated with interband transitions reach a plateaux as the light frequency $\Omega$ increases. 
For planar unstrained graphene, this plateaux is associated with a universal conductivity $\sigma_0=e^2/\hbar$ \cite{Peres2010}, so the role of the periodic curvature is to change this universal behavior along the deformed direction.
Indeed, in the limit of large incident frequencies, the effects due to temperature, chemical potential and the
periodic boundary condition are washed out. In the direction perpendicular to the deformation profile, called $y$ direction,
for sufficient large $\Omega$ one has ${\Re \,\sigma_{yy}^{\mathrm{inter}}}/{\sigma_0} \rightarrow 1$.
The high frequency limit for $\Re \sigma^{\mathrm{inter}}_{xx}$ has a plateaux whose values encodes
the geometry of the graphene layer.

\subsection{Expansion for small deformation limit}

Let us discuss the small deformation limit of Eq.(\ref{f012}) in more detail. 
By small deformation limit, one considers the case where
the arc length in units of $1 / \omega$ is essential unitary. If one assumes that $|h^2_x| \ll 1$, then one can approximate
\be
   \xi = 2 \pi  + \frac{ \langle h^2_x \rangle}{2} - \frac{ \langle h^4_x \rangle}{8} + \cdots
\ee
with $
   \langle h^n_x \rangle  = \int^{2 \pi}_{0} dv ~ h^n_x (v) \ $.

The function $F(n)$ can also be expanded in powers of $h^2_x$ and, after some straightforward algebra, 
the optical conductivity becomes
\bea
  \frac{\Re \sigma^{\mathrm{inter}}_{xx}  }{ \sigma_0 } & =&  1  - \frac{ \langle h^2_x \rangle}{2 \pi}
    + \frac{1}{16 \pi^2} \bigg( 3 \langle h^2_x \rangle^2 + 2 \pi  \langle h^4_x \rangle \bigg)  \nonumber \\
  & &
     + \frac{1}{8 \pi^2}   \sum_n \left| \int^{2 \pi}_{0} dv ~ e^{i n v } h^2_x(v) \right|^2 +
  \cdots
  \label{Eq:CondSmallDef}
\eea
It follows from Eq. (\ref{Eq:CondSmallDef}) that for small deformations of the graphene sheet, the optical conductivity is reduced
relative to the planar material value
and the optical transmittance is increased. Graphene becomes more transparent  for
large frequencies with deformation.

\begin{figure}[t] 
   \centering
   \includegraphics[scale=0.7]{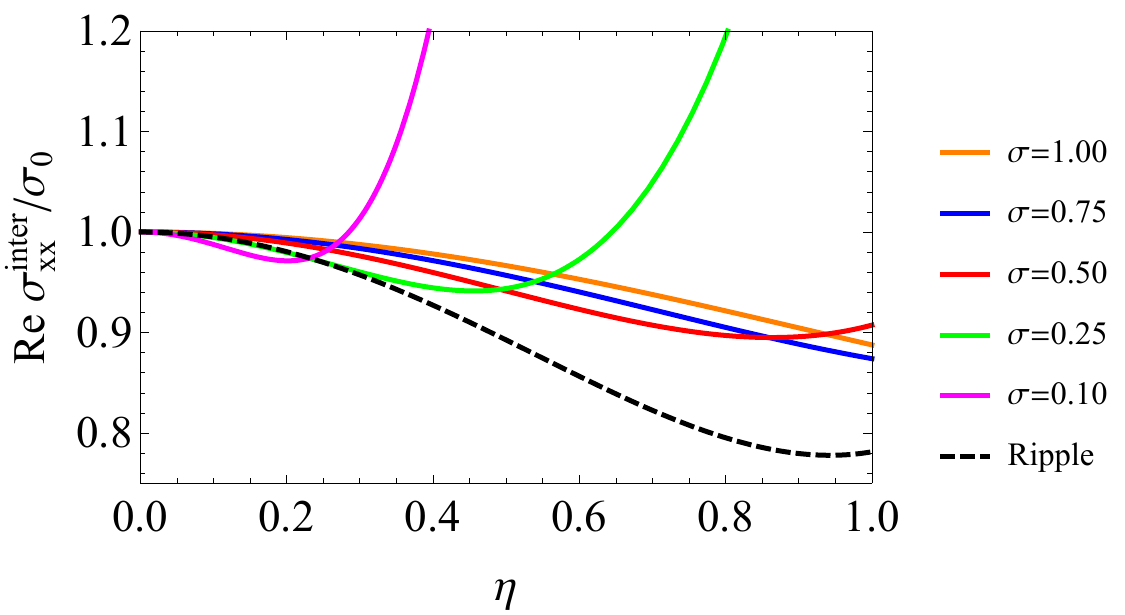}  \\
   \includegraphics[scale=0.7]{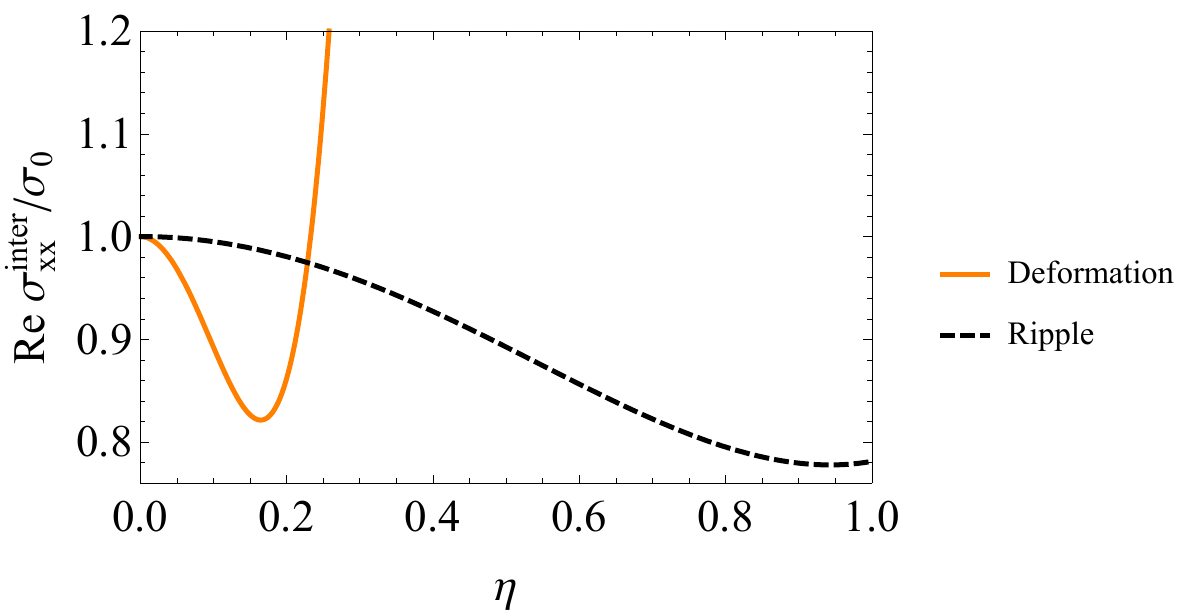}  \\ 
   \includegraphics[scale=0.7]{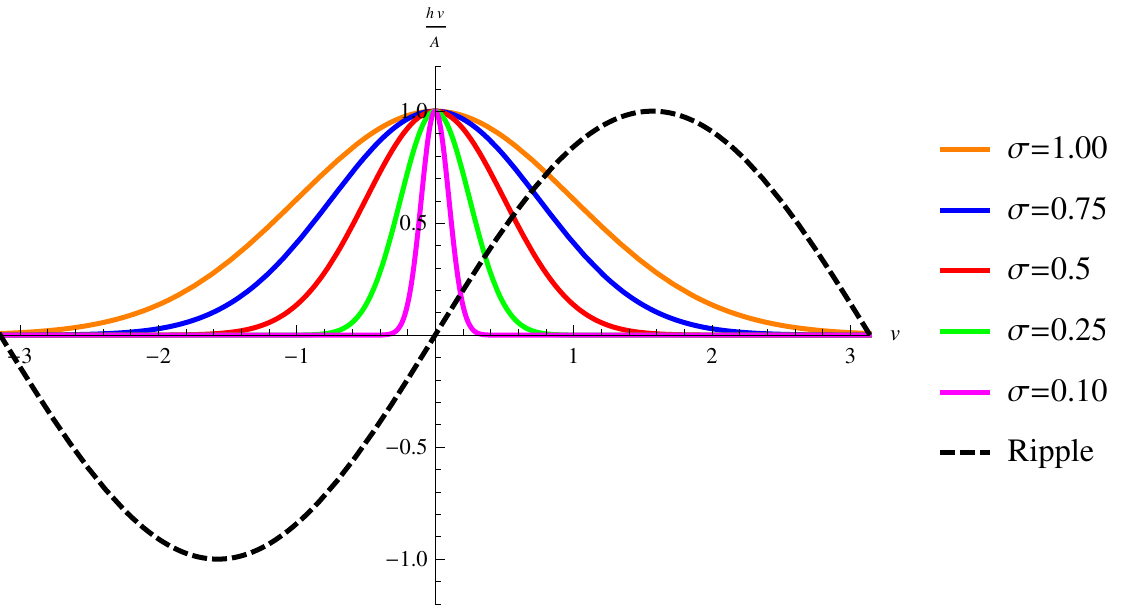}
   \caption{Optical conductivity (top), normalized to $ A = 1$, for  the ripple $h_1$ (\ref{h1}) and
   Gaussian deformations $h_2$ (\ref{h2}), with various widths.
 The bottom plot compares the optical conductivity for the deformations associated with $h_3$ (\ref{h3}) and $A_2 = 0$ and the ripple.}
   \label{fig:CondRippleGauss}
\end{figure}

Let us consider the following deformation profiles 
\bea
   h_1(x)  & = & A \, \sin ( \omega x ) \ , \label{h1}\\
   h_2 (x) & = & A \, \exp\left\{ - \frac{( \omega x - \pi )^2}{2 \sigma^2} \right\}  \ , \label{h2} \\
   h_3 (x) & = & A_1 \omega_1 x ( \omega_1 x - 2 \pi ) + A_2 \, \sin ( \omega_2 x ) \ . \label{h3}
\eea
in order to try to understand how deformations change the optical conductivity.
 The experimental results of~\cite{Ni2014}
will be analysed  using  these three different profiles.

The optical conductivity,  as given by expression (\ref{Eq:CondSmallDef}),
for the profiles $h_1$, $h_2$ (with $\eta = A \omega$) and $h_3$ (with $\eta = A _1 \omega_1$ and $A_2 = 0$)
can be seen on Fig.~\ref{fig:CondRippleGauss}. In what concerns the comparison between the ripple and the gaussian bump, it turns
out that the ripple has typically the smallest conductivity (highest transmittance). On the other hand, the deformation associated with
$h_3$ seems to be associated with an even smaller value of the conductivity when compared to the ripple profile. 
The rapid increase in the conductivity observed on Fig.~\ref{fig:CondRippleGauss}
for the $h_2$ and $h_3$ profiles is related to the validity of the approximation used in the calculation.

\section{Optical Transmittance \label{Sec:Trans}}

The transmittance is defined as the ratio between the outcoming and incoming light intensity and it is given by
$T = |T_x| \cos^2\theta + |T_y| \sin^2\theta$, where $\theta$ is the polarization angle, i.e. the angle between the
electric field component of light and the direction associated with the deformation $x$, and
\be
 |T_i| = \frac{1}{\left( 1 + \frac{\pi}{2} \, \alpha \,  \frac{\Re \sigma^{\mathrm{inter}}_{ii}}{\sigma_0}\right)^2} \approx
 1 - \pi \, \alpha \, \frac{\Re \sigma^{\mathrm{inter}}_{ii}}{\sigma_0}, \label{transmission}
\ee
to first order in the fine-structure constant $\alpha$. For small deformations one can write
\be
\frac{\Re \sigma^{\mathrm{inter}}_{ii}}{\sigma_0} = 1 + \delta_{ii},
\ee
and the transmittance becomes
\be
   T =  T_0 - \pi \, \alpha \,  \delta_{xx} \, \cos^2\theta,
   \label{Eq:Tsmalldef}
\ee
where $T_0 = 1 - \pi \, \alpha \approx 97.7 \%$ is the transmittance of pristine graphene~ \cite{Nair2008}. To first order in the deformation
it follows that
\be
   \delta_{xx} = - \frac{\langle h^2_x \rangle}{2 \pi} \ ,
\ee
and the geometric deformation increases  $T$  relative to the planar graphene.

\subsection{Unstrained graphene}

\begin{figure}[t] 
   \centering
   \includegraphics[scale=0.35]{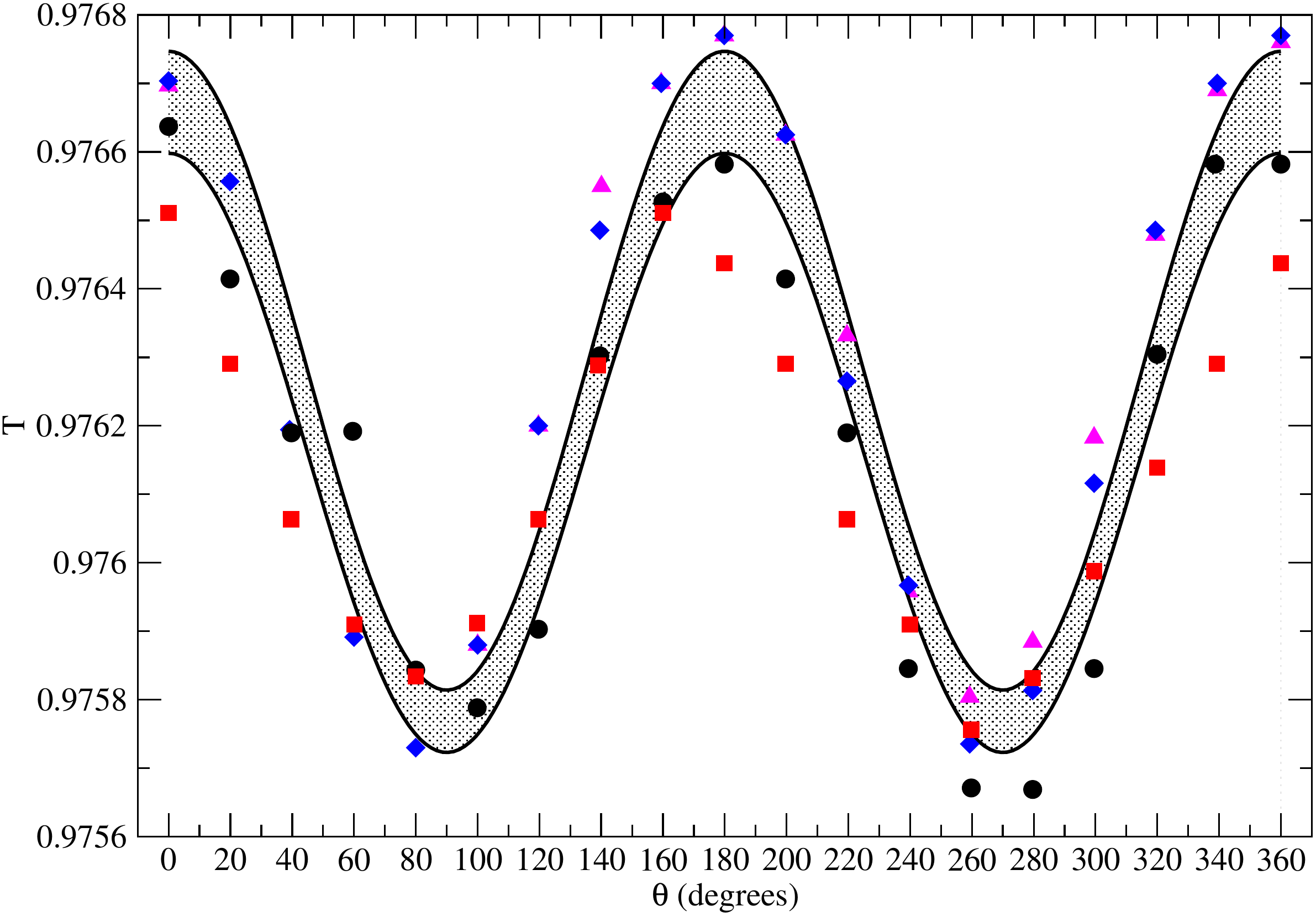}
   \caption{Optical transmittance of plane graphene. Experimental data extracted from Fig. 3 a) of ref.~ \cite{Ni2014},
      and are given by the box, triangle, diamond and circle, which represents different locations of the laser spot in the graphene. 
      The gray band includes all the four fittings with parameters given in Table \ref{table1}.}
   \label{Fig:Tplane}
\end{figure}

The optical transmittance for graphene and for deformed graphene was measured, as a function of the polarization angle $\theta$, 
in 
\cite{Ni2014}.
For plane graphene, the experimental values for $T$ at various locations of the graphene sheet can be seen on Fig.~\ref{Fig:Tplane}. The
experimental data for $T$ is consistent with the functional form (\ref{Eq:Tsmalldef}). For each data set in Fig.~\ref{Fig:Tplane}, a fit to $a + b
\cos^2\theta$ gives the parameters presented in Table \ref{table1}, where the parameters are 
labelled by the color of the corresponding data set used in the fit.

\begin{table}[]
\centering
\caption{Transmittance parameters  for plane graphene.}
  \label{table1}
   \begin{tabular}{l@{\hspace{0.5cm}}l@{\hspace{0.3cm}}l@{\hspace{0.3cm}}l}
                                  &  $a$                    &    $b$                   & $\chi^2/d.o.f$ \\
   black            &   0.975723(37)    &   0.000875(59)    &  $9 \times 10^{-9} \ ,$ \\
   red               &  0.975807(33)     &   0.000621(52)    &  $7 \times 10^{-9} \ ,$ \\
   blue             &  0.975779(42)     &   0.000979(65)    &  $1 \times 10^{-8} \ ,$  \\
   magenta      &  0.975814(46)     &   0.000933(72)    &  $1 \times 10^{-8} \ .$
\end{tabular}
\end{table}

In our analysis the quoted errors on the coefficients $a$ and $b$ are associated purely to the fit.
The values of the $\chi^2/d.o.f.$ are quite small but seem to support that $T$ is well described by Eq. (\ref{Eq:Tsmalldef}). We
call reader attention that the value for $a \approx 0.976$ is quite close to the theoretical estimate $T_0 = 0.977$.

From Fig. 1 of ref.~\cite{Ni2014}  a very crude estimate of the parameter $b$ can be made. Indeed, reading from the figure and assuming a
ripple shape it follows that $A \sim 90 $ nm and $\lambda \sim 25$ $\mu$m giving $\eta_{exp} \sim 0.2$.
The results of the fits using the ripple  profile (\ref{h1}) give $\eta = 0.28$ (black), 0.24 (red), 0.29 (magenta) and 0.30 (blue),
which are in the same ballpark as the very crude estimate obtained directly from the figure of the graphene slab~\cite{Ni2014}.

\begin{figure}[t] 
   \centering
   \includegraphics[scale=0.5]{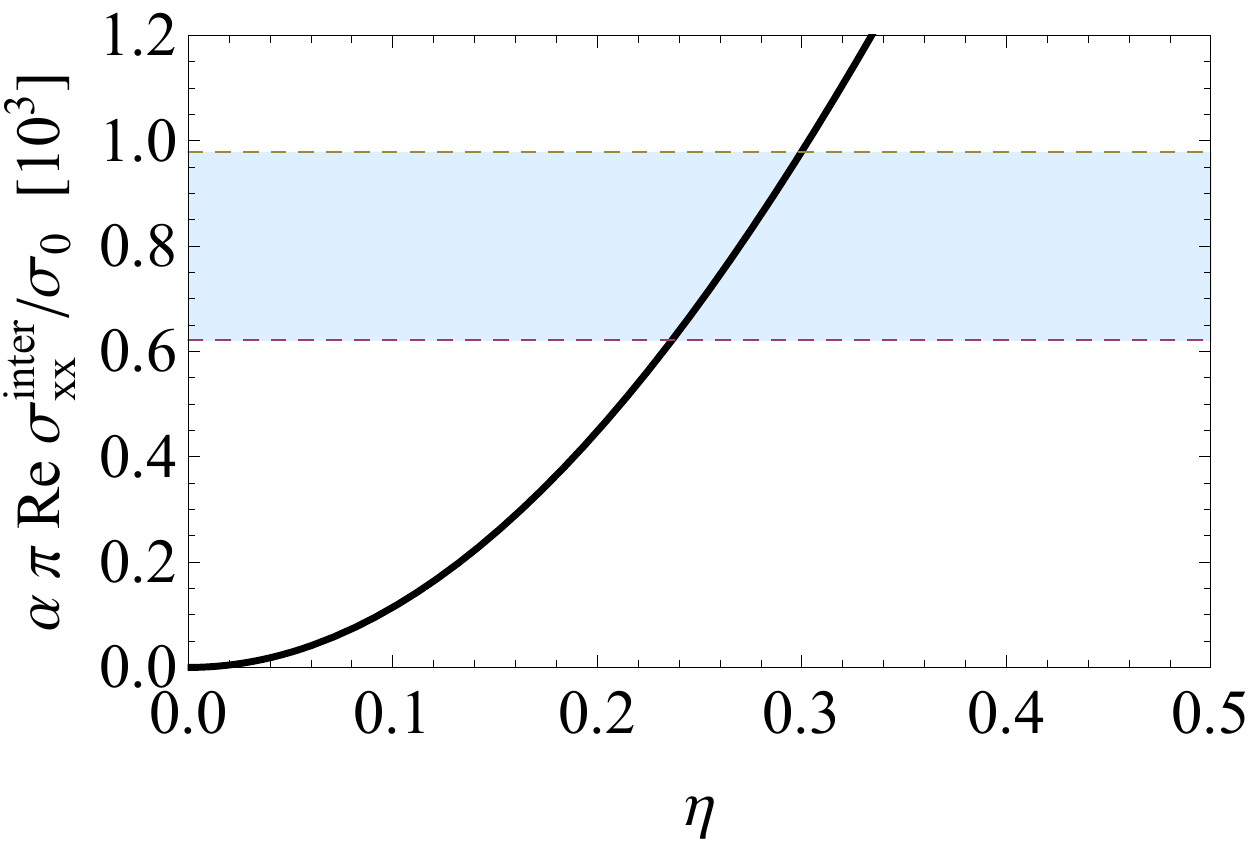} \\
   \includegraphics[scale=0.5]{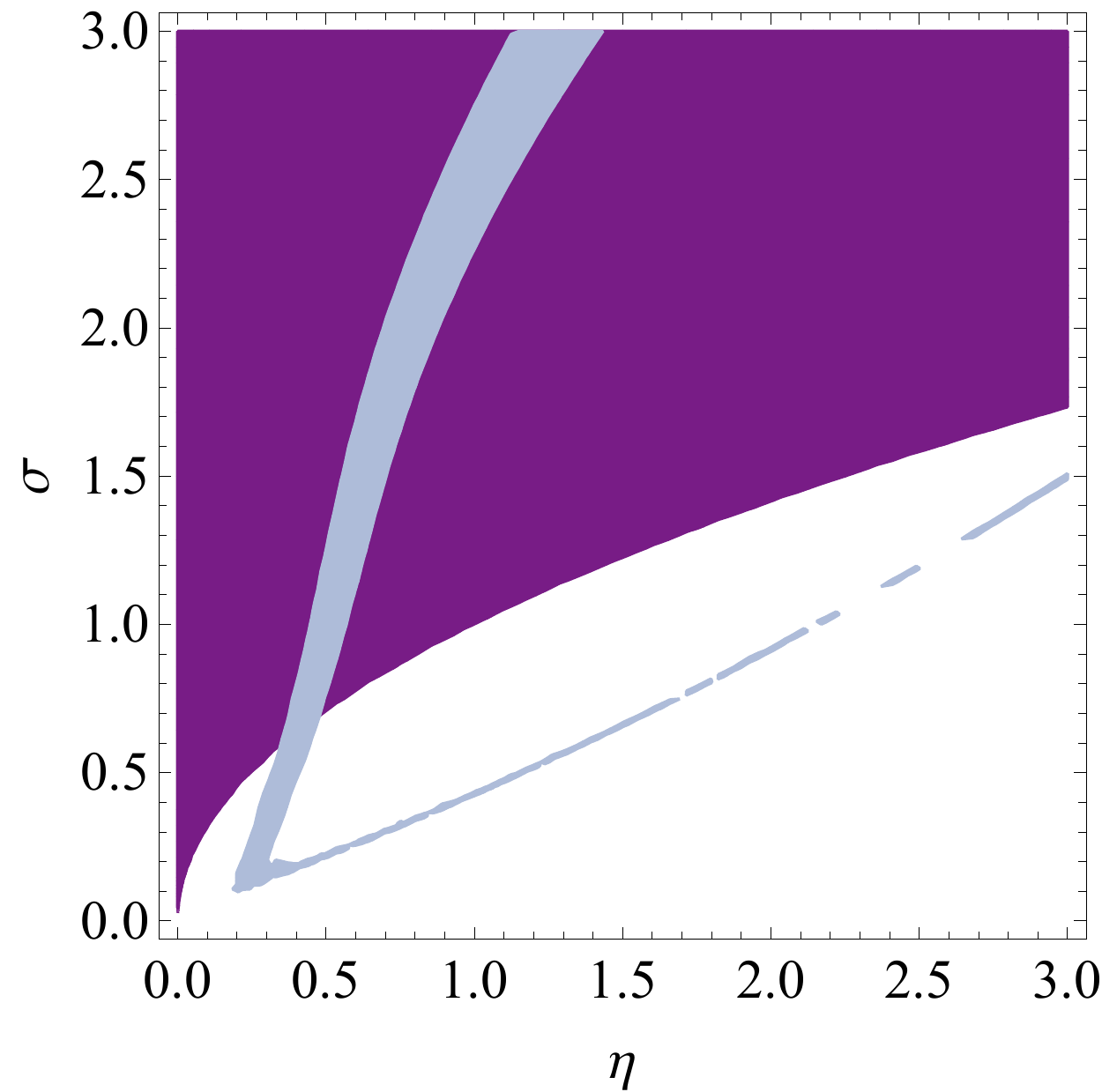} \\
   \includegraphics[scale=0.5]{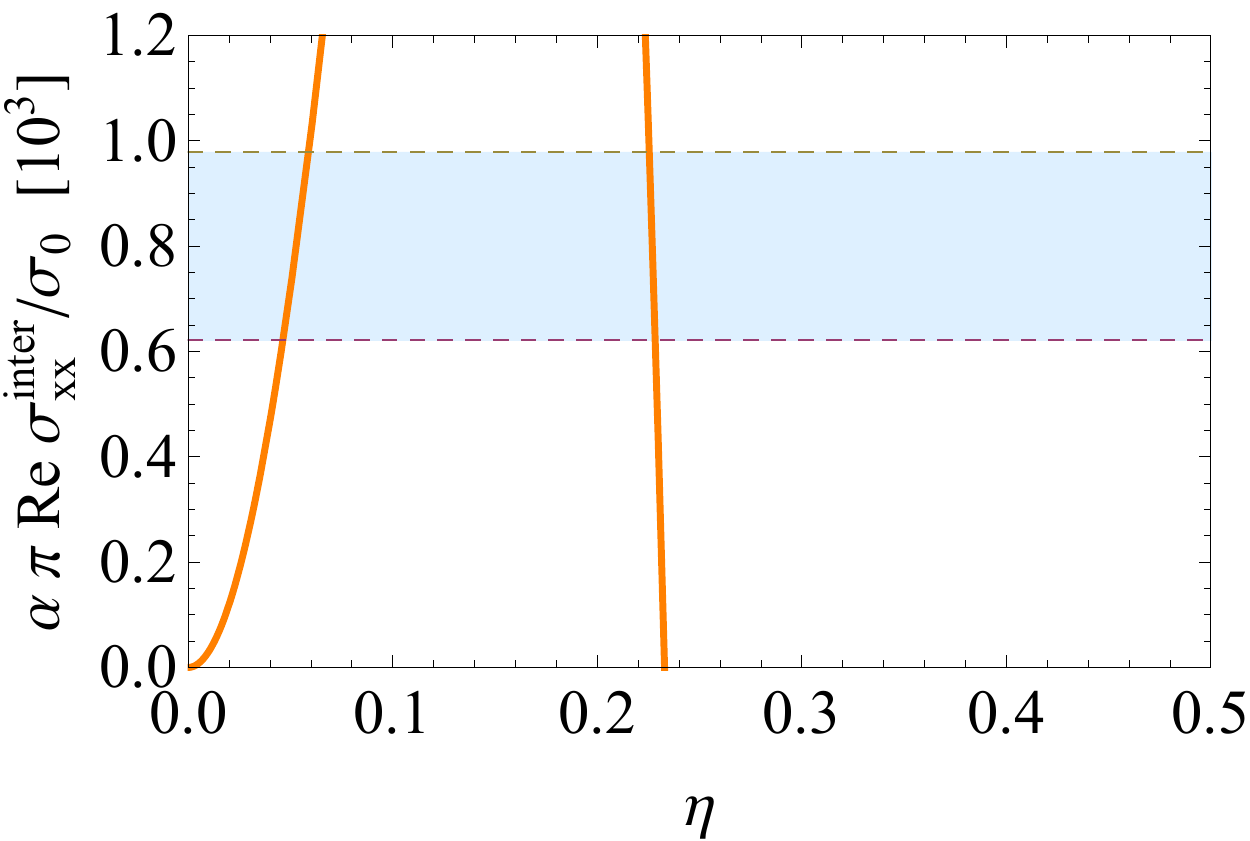}
   \caption{Optical transmittance of unstrained graphene.
     The top and bottom frames show the allowed region of parameters for the ripple, $h_1$,
     and for the deformation $h_3$ with $A_2 = 0$, respectively. The blue region shows
   the set of parameters compatible with the outcome of the fits discussed in the text.
    The middle frame shows the allowed range of  $\eta$ and $\sigma$ values for the Gaussian profile, $h_2$,
    consistent with the fitted data sets. The blue area represents
   the constraints from the fits, while the magenta region is defined by requiring the validity of the small deformation expansion which
   requires $\eta/\sigma^2 < 1$. }
   \label{Fig:Allowed}
\end{figure}

In the analysis we assumed that the deformation of graphene is a ripple structure and therefore we only considered so far the 
profile $h_1$. However, the experimental data is also compatible with the profiles $h_2$ and $h_3$.

On Fig.~\ref{Fig:Allowed}, we show the allowed range of parameters for the various profiles compatible with the interval of values
coming out from the fits to the experimental transmittance reported in Fig.~\ref{Fig:Tplane}.
As we have already discussed, the ripple profile $h_1$ suggests a value for $\eta$ that is compatible with the crude estimate from 
the actual graphene sample giving an $\eta_{exp} \sim 0.2$ (see upper frame in Fig.~\ref{Fig:Tplane}). 

The allowed range of fitting parameters of the Gaussian profile $h_2$ is shown in the middle frame of Fig.~\ref{Fig:Tplane} and it seems to
favour $\eta \ge 0.5$, if one assumes small deformations of graphene. Such value for $\eta $ is hardly consistent with
$\eta_{exp}$. On the other hand, the quadratic profile $h_3$ with $A_2=0$ has a parametric region consistent with $\eta_{exp}$, 
as shown in the bottom frame of the figure. 

From the analysis of the various profiles and relying on the comparison between the allowed values of $\eta$ and $\eta_{exp}$, see Table~\ref{table1},
one concludes that experimental data favours the ripple and the quadratic structures.

\subsection{Uniaxial strained graphene}

\begin{figure}[t] 
   \centering
   \includegraphics[scale=0.31]{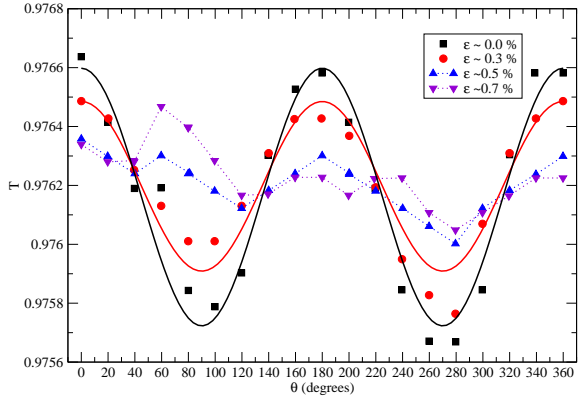}
   \caption{Transmittance for bended graphene layer. Experimental data from Fig. 3-b of ref.~\cite{Ni2014},
   where $\varepsilon$ is the strain fraction along the bending direction. The dotted lines guide the eyes through the
   experimental data with $\varepsilon=0.5\%$ and $0.7\%$.
   The solid curves are our fittings to the data with  $\varepsilon=0$ (black line - large amplitude) and
   $\varepsilon=0.3\%$ (red line - small amplitude).}
   \label{fig:T_defor_exp}
\end{figure}

In ref.~\cite{Ni2014} the graphene optical transmittance was measured both for unstrained and strained material. The data for 
this later case were obtained by bending the graphene layer and can be seen on Fig.~\ref{fig:T_defor_exp}.
The isotropy of the hexagonal lattice elasticity allows to define the strain tensor to lowest order in terms of
the uniaxial strain $\varepsilon$, the associated bending direction with respect to the zig-zag direction,
and the Poisson ratio. The experiments reported in ref.~\cite{Ni2014} were performed for $\varepsilon = 0.3 \%$, $0.5 \%$ and $0.7 \%$. 
The model based on Quantum Field Theory in curved space is valid for small deformations and 
we focus on flat graphene and on the deformation for $\varepsilon = 0.3 \%$.

We would like to mention that the effects of in-plane deformations associated to strain and out-of-the plane deformations on the optical conductivity
of graphene-like materials, within the formalism of a Dirac equation in curved space, were explored in ref.~\cite{PaulaFBS15}. 
In our analysis of the experimental results, the contributions of in-plane deformations and of possible changes in the Fermi velocity of the electrons
are
not taken into account.

\begin{figure}[t]
   \centering
   \includegraphics[scale=0.55]{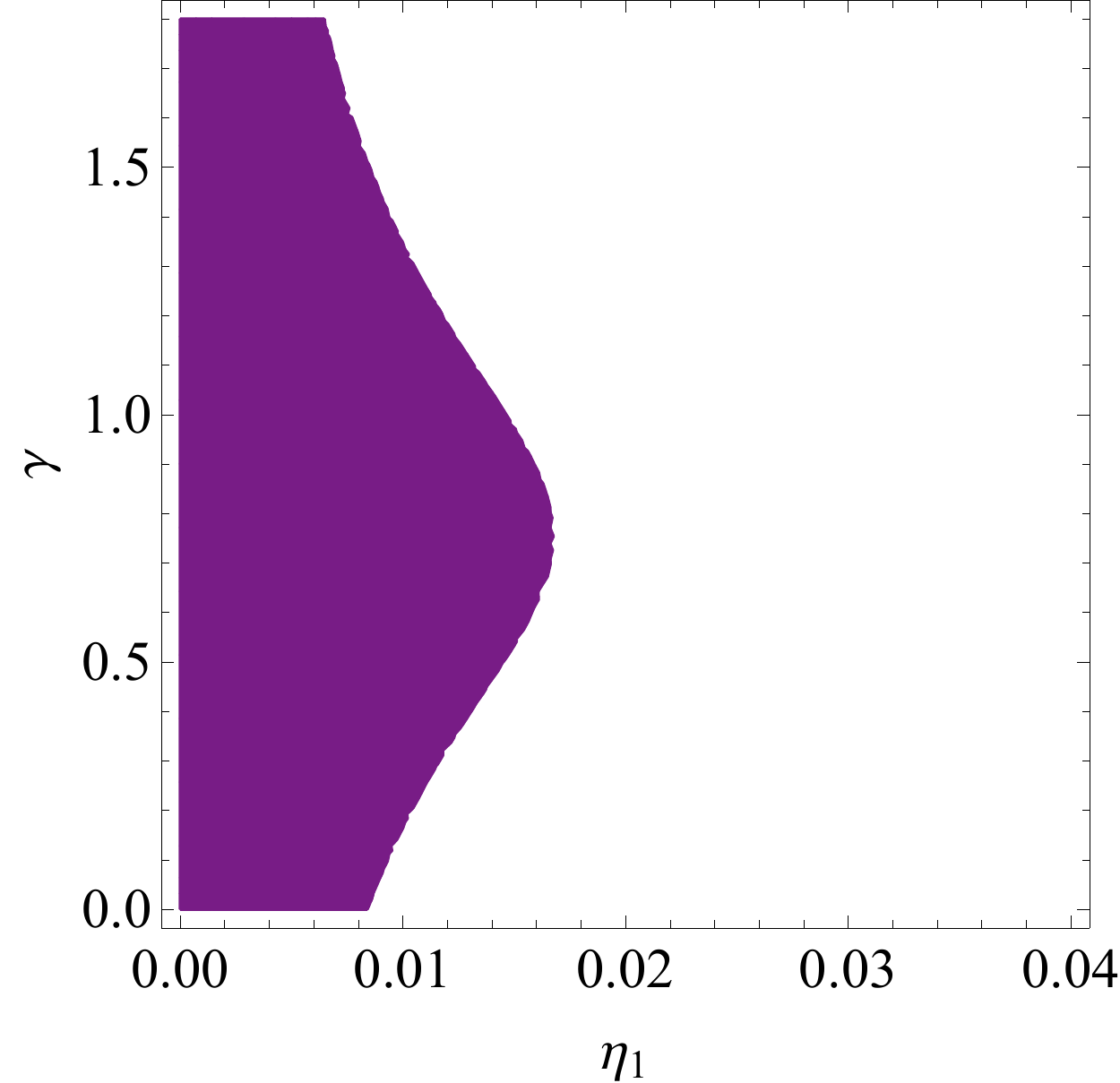}
   \caption{Allowed region of parameters for the quadratic term in the
   $h_3$ profile compatible with the fitted data for $\varepsilon = 0.3\%$.}
   \label{fig:example}
\end{figure}

The data characterized by $\varepsilon = 0.3 \%$ follows the same pattern as observed for unstrained  graphene, 
with the transmittance being compatible with a $\cos^2\theta$ dependence with the polarization angle.
The data shows a decrease on the oscillation amplitude with respect to the unstrained case, and  the fitted value of $b$ 
is somewhat smaller than the ones presented in Table \ref{table1}. Indeed, the fit to the data gives 
$a = 0.975909(30)$, $b = 0.000576(47)$ for a $\chi^2/d.o.f. = 6 \times 10^{-9}$.
In terms of $b$, the deformation implies a reduction of its value by about 50\% when compared to the unstrained case.
This can be accommodated within the profile $h_3$, which combines the ripple structure considered previously with a smooth quadratic type of deformation,
if one takes $\eta_2 = A_2 \omega_2$ slightly smaller than
the corresponding outcome for the red points in unstrained graphene ($\eta = 0.24$) and assumes that the quadratic part of the profile 
$h_3$ simulates the bending of the graphene layer.
Fig.~\ref{fig:example} shows the allowed parameter range compatible with $b = 0.0005$ and a 10\% relative error;
in the figure $\gamma = \omega_1 / \omega_2$.
The data clearly favours small values of $\eta_1=A_1\omega_1\,\,(< 0.02)$, but leave $\omega_1$ essentially unbounded.
This allows us to identify $\omega_1$ with the curvature induced by the bending of the graphene layer.

The range of estimated parameters ($\eta$, $\gamma$, $\omega_1$) allowed from the analysis of the data for the strained graphene 
can be accommodated within our theoretical description of a strained graphene layer. Further, the estimated values are also compatible
with those provided from the analysis of the unstrained case. 

\section{Conclusions}
The consistency between the quantitative and qualitative analysis of the light transmittance observed for
unstrained graphene and the quantum field theory framework in curved space suggests that the light
transmittance is determined by the ripple structure of the graphene sheet.
For strained graphene and small enough deformations, the experimental data is well reproduced by a geometrical
deformation of the graphene sheet which takes into account the ripple structure, as observed in the unstrained case,
together with a smooth deformation of the original material, parametrized in our case by a quadratic function which
adds to the sinusoidal ripples.

The analysis performed here shows once more that the field theoretical framework in curved spaces can be a useful tool in the
investigation of electronic and optical properties of graphene or, from another point of view, offers the possibility of engineering graphene 
to study quantum mechanics in deformed space-times at tabletop experiments.

\acknowledgments
The authors thanks Guang-Xin Ni for calling our attention to the experiment \cite{Ni2014} and for helpful
discussions. The authors also thanks V Pereira for useful discussions.
The authors acknowledge financial support from the Brazilian
agencies FAPESP (Funda\c c\~ao de Amparo \`a Pesquisa do Estado de
S\~ao Paulo) and CNPq (Conselho Nacional de Desenvolvimento
Cient\'ifico e Tecnol\'ogico). OO acknowledges financial support from grant 2014/08388-0 from S\~ao Paulo Research Foundation (FAPESP).


\begin{thebibliography}{0}

\bibitem{Novoselov2004}
K. S. Novoselov, A. K. Geim, S. V. Morozov, D. Jiang, Y. Zhang, S. V. Dubonos, I. V. Grigorieva, A. A. Firsov,
Science \textbf{306} (2004) 666 .
\bibitem{Novoselov2005}
K. S. Novoselov, D. Jiang, T. Booth, V. V. Khotkevich, S. M. Morozov, A. K. Geim, 2005, Proc. Natl. Acad. Sci.
U.S.A. \textbf{102} (2005) 10451.
\bibitem{Neto2009}
A. H. Castro Neto, F. Guinea, N. M. R. Peres, K. S. Novoselov, A. K. Geim, Rev. Mod. Phys. \textbf{81} (2009) 109.
\bibitem{Vozmediano2010}
M. A. H. Vozmediano, M. I. Katsnelson and F. Guinea, Phys. Rep. \textbf{496} (2010) 109.
\bibitem{Peres2010}
N. M. R. Peres, Rev. Mod. Phys. \textbf{82} (2010) 2673.
\bibitem{Chaves2011}
A. Chaves, G. Lima, W. de Paula, C. Cordeiro, A. Delfino, T. Frederico, O. Oliveira, Phys. Rev. B \textbf{83} (2011) 153405.
\bibitem{Oliveira2011}
O. Oliveira, C. E. Cordeiro, A. Delfino, W. de Paula, T. Frederico, Phys. Rev. B \textbf{83}  (2011) 155419.
\bibitem{Popovici2012}
C. Popovici, O. Oliveira, W. de Paula, T. Frederico, Phys. Rev. B \textbf{85} (2012) 235424.
\bibitem{Cordeiro2013}
C. E. Cordeiro, A. Delfino, T. Frederico, O. Oliveira, W. de Paula, Phys. Rev. B \textbf{87} (2013) 045429.
\bibitem{Morozov2006}
S. V. Morozov, K. S. Novoselov, M. I. Katsnelson, F. Schedin, L. A. Ponomarenko, D. Jiang, A. K. Geim,
Phys. Rev. Lett. \textbf{97} (2006) 016801.
\bibitem{Meyer2007}
J. C. Meyer, A. K. Geim, M. I. Katsnelson, K. S. Novoselov, T. J. Booth, S. Roth, Nature \textbf{446} (2007) 60.
\bibitem{Ishigami2007}
M. Ishigami, J.H. Chen, W.G. Cullen, M.S. Fuhrer, E.D. Williams, Nano Letters \textbf{7} (2007) 1643.
\bibitem{Stolyarova2007}
E. Stolyarova, K.T. Rim, S. Ryu, J. Maultzsch, P. Kim, L.E. Brus, T.F. Heinz, M.S. Hybertsen, G.W. Flynn,
PNAS \textbf{104} (2007) 9209.
\bibitem{Tapaszto2012}
L. Tapaszto, T. Dumitrica, S. J. Kim, Sung, P. Nemes-Incze, C. Hwang, L. P. Biro, Laszlo,
Nat. Phys. \textbf{8} (2012) 739.
\bibitem{Fasolino2007}
A. Fasolino, J. H. Los, M. I. Katsnelson, Nature Mater. \textbf{6} (2007) 858.
\bibitem{GuineaNature2010}
F. Guinea, M. I. Katsnelson, A. K. Geim, Nature Phys. \textbf{6} (2010) 30.
\bibitem{Levy2010}
N. Levy, S. A. Burke, K. L. Meaker, M. Panlasigui, A. Zettl, F. Guinea, A. H. Castro Neto, M. F. Crommie,
Science \textbf{329} (2010) 544.
\bibitem{Juan2011}
F. de Juan, A. Cortijo, M. A. H. Vozmediano, A. Cano, Nature Phys. \textbf{7} (2011) 810.
\bibitem{Chaves2014}
A. J. Chaves, T. Frederico, O. Oliveira, W. de Paula, M. C. Santos, J Physics: Condensed Matter \textbf{26} (2014) 185301.
\bibitem{Jakubsky2014}
V. Jakubsk\'y, D. Krej\v{c}i\v{r}\'{ \i}k, Ann. Phys. \textbf{349} (2014) 268.
\bibitem{Juan2012}
F. de Juan, M. Sturla and M. A. H. Vozmediano, Phys. Rev. Lett. \textbf{108} (2012) 227205.
\bibitem{Pereira2009}
V. M. Pereira, A. H. Castro Neto, N. M. R. Peres, Phys. Rev. \textbf{B 80} (2009) 045401.
\bibitem{Winkler2010}
R. Winkler, U. Z\"ulicke, Phys. Rev. \textbf{B 82} (2010) 245313.
\bibitem{Pellegrino2011}
F. M. D. Pellegrino, G. G. N. Angilella, R. Pucci, Phys. Rev. \textbf{B 84} (2011) 195404.
\bibitem{Linnik2012}
T. L. Linnik, J. Phys. Condens. Matter \textbf{24} (2012) 205302.
\bibitem{Lu2012}
J. Lu, A. H. Castro Neto, K. P. Loh, Nat. Comm. \textbf{3} (2012) 823.
\bibitem{Vozmediano2008}
F. Guinea, M. I. Katsnelson and M. A. H. Vozmediano, Phys. Rev. B {\bf 77} (2008) 075422.
\bibitem{Guinea2008}
F. Guinea, B. Horovitz, P. Le Doussal, Phys. Rev. \textbf{B 77} (2008) 205421.
\bibitem{Isacsson2008}
A. Isacsson, L. M. Jonsson, J. M. Kinaret, M. Jonson, Phys. Rev. \textbf{B 77} (2008) 035423.
\bibitem{Pereira2009prl}
V. M. Pereira, A. H. Castro Neto, Phys. Rev. Lett. \textbf{103} (2009) 046801.
\bibitem{Guinea2010}
F. Guinea, A. K. Geim, M. I. Katsnelson, K. S. Novoselov, Phys. Rev. B \textbf{81} (2010) 035408.
\bibitem{Pereira2010}
V. M. Pereira, R. M. Ribeiro, N. M. R. Peres, A. H. Castro Neto, EPL \textbf{92} (2010) 67001.
\bibitem{Pellegrino2010}
F. M. D. Pellegrino, G. G. N. Angilella, R. Pucci, Phys. Rev. B \textbf{81} (2010) 035411.
\bibitem{Zhang2011}
D.-B. Zhang, E. Akatyeva, and T. Dumitric\v{a}, Phys. Rev. Lett. \textbf{106} (2011) 255503.
\bibitem{Kerner2012}
R. Kerner, G. G. Naumis, W. A. G\'omez-Arias, Physica B \textbf{407} (2012) 2002.
\bibitem{Yan2013}
H. Yan, Z.-D. Chu, W. Yan, M. Liu, L. Meng, M. Yang, Y. Fan, J. Wang, R.-F. Dou, Y. Zhang, Z. Liu, J.-C. Nie, L. He,
Phys. Rev. \textbf{B 87} (2013) 075405.
\bibitem{OlivaLeyva2014}
M. Oliva-Leyva, G. G Naumis, J Physics: Condens. Matter \textbf{26}, 125302 (2014); \textit{Corrigendum}: J. Phys.: Condens. Matter 26 (2014) 279501.
\bibitem{Nguyen2014}
M. C. Nguyen, V. H. Nguyen, H.-V. Nguyen, P. Dollfus, Semicond. Sci. Technol. \textbf{29} (2014) 115024.
\bibitem{Scniepp2008}
H. C. Schniepp, K. N. Kudin, J.-L. Li, R. K. Prudhomme, R. Car, D. A. Saville, I. A. Aksay
ACS Nano, \textbf{2} (12) (2008) 2577.
\bibitem{Xu2009}
K. Xu, P. Cao, and J. R. Heath, Nano Lett. \textbf{9} (2009) 4446.
\bibitem{Bao2009}
W. Bao, F. Miao, Z. Chen, H. Zhang, W. Jang, C. Dames, C. N. Lau, Nature Nanotechnology {\bf 4} (2009) 562.
\bibitem{Wang2011}
Y. Wang, R. Yang, Z. Shi, L. Zhang, D. Shi, E. Wang, G. Zhang, ACS Nano, {\bf 5} (5) (2011) 3645.
\bibitem{Koch2012}
M. Koch, F. Ample, C. Joachim, L. Grill, Nature nano. \textbf{7} (2012) 713.
\bibitem{Bai2014}
K.-K. Bai, Y. Zhou, H. Zheng, L. Meng, H. Peng, Z. Liu, J.-C. Nie, L. He, arXiv:1404.4407
\bibitem{Ni2014}
G.-X. Ni, H.-Z. Yang, W. Ji, S.-J. Baeck, C.-T. Toh, J.-H. Ahn, V. M. Pereira, B. \"Ozyilmaz, Adv. Mater., \textbf{26} (2014) 1081.
\bibitem{Park2014}
Y. J. Park, S.-K. Lee, M.-S. Kim, H. Kim, J.-H. Ahn, ACS Nano \textbf{8} (2014) 7655.
\bibitem{Plantey2014}
A. Reserbat-Plantey, D. Kalita, Z. Han, L. Ferlazzo, S. Autier-Laurent, K. Komatsu, C. Li, R. Weil, A. Ralko,  L. Marty,
S. Gu\'eron, N. Bendiab, H. Bouchiat, V. Bouchiat, Nano Lett. \textbf{14} (2014) 5044.
\bibitem{Dong2014}
B. Dong, P. Wang, Z.-B. Liu, X.-D. Chen, W.-S. Jiang, W. Xin, F. Xing, J.-G. Tian, Nanotechnology \textbf{25} (2014) 455707.
\bibitem{Casillas2014}
G. Casillas, U. Santiago, H. Barrón, D. Alducin, A. Ponce and M. Jos\' e-Yacam\'an, Journ. of Phys. Chem. C \textbf{119} (2014) 710.
\bibitem{Nair2008}
R. R. Nair, P. Blake, A. N. Grigorenko, K. S. Novoselov, T. J. Booth, T. Stauber, N. M. R. Peres, A. K. Geim,
Science \textbf{320} (2008) 5881.
\bibitem{PaulaFBS15} W. de Paula , A. J. Chaves, O. Oliveira, T. Frederico, Few-Body Systems \textbf{56} (11-12) (2015) 915.
 \end{thebibliography}
\end{document}